\documentstyle[12pt,worldsci]{article}
\newcommand{\pp}{$\pi\pi$ }
\newcommand{\kk}{$K\overline{K}$ }
\newcommand{\roro}{$\sigma\sigma$ }
\newcommand {\eq}{\begin{equation}}
\newcommand {\qe}{\end{equation}}
\newcommand {\la}{\lambda}
\newcommand{\fo}{$f_0(980)$ }
\newcommand{\epw}{$f_0(1400)$ }
\newcommand{\epsig}{$f_0(500)$ }

\input psfig

\def\xslide#1#2#3#4#5#6#7{\centerline{\psfig
{figure=#1,height=#2,bbllx=#3bp,bblly=#4bp,bburx=#5bp,bbury=#6bp,width=#7,clip=}}}

\begin{document}
\title{PHENOMENOLOGY OF SCALAR MESONS}
\author{Leonard Le\'sniak\\
{\em H. Niewodnicza\'nski Institute of Nuclear Physics, \\PL 31-342 Krak\'ow,
 Poland}}
 
\vspace{0.3cm}

\maketitle

\setlength{\baselineskip}{2.6ex}

\vspace{0.7cm}

\begin{abstract}

 A separable potential
model of three coupled channels $\pi \pi$, \kk and an effective $2\pi
2\pi$ has been constructed and applied in the analysis of the
isoscalar S-wave \pp and \kk phase shifts and inelasticities. A relatively 
narrow scalar resonance of mass 1400-1460 MeV has been found. This resonance is 
consistent with the recent CERN observation of an eventual scalar glueball at 
1500 MeV. Other resonances at lower energies like a wide $\sigma$ meson and 
a narrow \fo have been confirmed.
 
\end{abstract}

\vspace{0.7cm}

    Scalar meson spectroscopy is full of open questions. The internal structure
of scalars is also very controversial, for example the $f_0(980)$ state
is interpreted as $q \bar q,q \bar qq \bar q$ or \kk quasibound
state. Mixing of scalar mesons with scalar glueballs is possible; for example 
the $f_0(1500)$ resonance, observed by the Crystal Barrel Collaboration in 
$\bar p p$
annihilation at CERN, was interpreted as a candidate for the lowest scalar
glueball \cite{amslerklkl}. The phenomenology of scalar mesons must include studies
of interactions between different mesons produced in the decay channels of 
scalar mesons. 
In Ref. \cite{klm} the scalar meson spectrum was studied in terms of
a relativistic \pp and \kk coupled channel 
model from the $\pi\pi$ threshold up to 1400 MeV. 
The phenomenological parameters were constrained by fitting the
$S$-wave data extracted from the experimental cross sections on the 
$\pi^+\pi^-$ production by $\pi^-$ scattering on unpolarized hydrogen target 
\cite{grayer} and further constraints were
imposed by the \kk phase shift analysis of Ref. \cite{cohen}~.

Recently, authors of Ref. \cite{klr} have analysed data obtained on
a polarized target by the CERN-Cracow-Munich group for the
$\pi^-p_{\uparrow} \to \pi^+\pi^-n$ reaction. Separation
of the $\pi$ and $a_1$ exchange amplitudes in this reaction was
then possible for the first time. From a set of four solutions for
the isoscalar $S$-wave phase shifts up to 1600 MeV, two of them
("down-flat" and "up-flat") satisfy the unitarity constraint. The
"down-flat" solution is in good agreement with the former solution
of Ref. \cite{grayer} up to 1400 MeV. Above 1400 MeV one observes an
increase of the phase shifts and larger inelasticities than those seen
in ~\cite{grayer}~. 
This could be a manifestation of the presence of 
scalar mesons $f_0(1370)$ or $f_0(1500)$ in that energy range.
There a strong four--pion production has been
observed in different experiments together with an
evidence of clustering into $\sigma\sigma$ or $\rho\rho$ pairs.

 In Ref. \cite{kll} we have extended  the isospin 0 $S$--wave relativistic 
2--channel model of Ref. \cite{klm} by adding to its $\pi\pi$ and \kk
channels an effective third coupled channel, here called $\sigma\sigma$.
The reaction amplitudes $T$ satisfy a system of the coupled channel 
Lippmann-Schwinger equations:

\begin{equation} 
T=V + VGT ,
\label{t}
\end {equation} 
where $V$, $G$ and $T$ are $3 \times 3$ matrices, $V$ is the interaction matrix 
and $G$ is the diagonal matrix of channel propagators:
\eq 
G_j(E,k_j)=\frac{1}{E-2E_j(k_j)+i\epsilon},\ \ \  \epsilon \to 0(+), \ \ \ j=1,2,3. 
\label{pro}\qe
In  Eq. (\ref{pro}) $E$ is the total energy, $E_j= \sqrt{m_j^2+{k_j}^2}$ and $m_j$ 
is the meson mass in channel $j$.  We consider meson pairs of same 
mass and momentum $k_j$ in their centre of mass system. In this way 
9 meson-meson reactions are simultaneously described  by Eq.(\ref{t}):
\begin{equation} \left(
\begin{array}{ccc}
\pi\pi \rightarrow \pi\pi & \pi\pi \rightarrow K\overline{K} & 
\pi\pi \rightarrow \sigma\sigma  \\
K\overline{K} \rightarrow \pi\pi & K\overline{K} \rightarrow K\overline{K} &
K\overline{K} \rightarrow \sigma\sigma  \\
\sigma\sigma \rightarrow \pi\pi & \sigma\sigma\rightarrow K\overline{K} &
\sigma\sigma \rightarrow \sigma\sigma \\
\end{array} \right).
\end {equation} 

We choose a separable form of the interaction:
\eq <p|V_{ij}|q> = \sum_{\alpha=1}^{n}\la_{ij,\alpha}\ 
                             g_{\alpha,\,i}(p)\ g_{\alpha,\,j}(q),                  
                              \ \ i\, , j = 1, 2, 3,
                              \label{pot}\qe
where $\la_{i\,j,\alpha}$ are coupling constants and 
\eq  g_{\alpha,\,j}(p)= \sqrt{\frac{4\pi}{m_j}}
                             \frac{1}{p^2+(\beta_{\alpha,\,j})^2}
                             \label{for}\qe
are form factors which depend on the relative centre of mass meson 
momenta $p$ in the final channel or $q$ in the initial channel.  In the
\pp channel ($j=1$) we choose a rank-2 separable potential ($n=2$) and in
the other channels a rank-1 potential ($n=1$).
 Altogether, this model has 13
parameters:  9 coupling constants $\la_{i\,j,\alpha}$ and 4 range
parameters $\beta_{\alpha,\,j}$. 

The model is unitary. The diagonal S- matrix elements are parametrized as
\eq S_{jj} = \eta_je^{2i\delta_j}, ~~j = 1,2,3,\qe
where $\eta_j$ and $\delta_j$ are channel $j$ inelasticities and 
phase shifts, respectively. Expressions for nondiagonal elements can be
found in Ref. \cite{lles96}~. Some of the $S$-matrix poles in the
complex energy plane can be interpreted as resonances. 
We fit the existing experimental results on the
$\pi\pi$ $S$-wave isoscalar phase shifts together with inelasticity in
the $\pi\pi$ channel and with the \kk phase shifts. 

Below 600 MeV we have used data from the $K_{e4}$ decay and above 600 MeV the 
"down-flat" and "up-flat" solutions of the analysis of Ref. \cite{klr} on a
polarized target. In addition we have used the results of the analysis of 
reactions $\pi^-p \to K^+K^-n$ and $\pi^+n \to K^-K^+p$ , although
targets were unpolarized there \cite{cohen}~. Four fits for $\delta_{\pi\pi}$ and 
$\eta_{\pi\pi}$ are compared to the experiment in Fig. 1.
Above 1400 MeV both "down-flat" and
"up-flat" data indicate a decrease towards small values of $\eta\ 
(\eta \approx 0.6$ to $0.7)$, albeit with large errors. One can notice 
a better agreement with data of the 3--channel model in comparison with the 
2--channel one.
The main difference between the 2-- and 3--channel fits lies in $\eta$
above 1400 MeV, where the opening of the 
$\sigma\sigma$ channel
leads to a fast decrease of inelasticity parameters. In order to achieve this behaviour, 
couplings between the $\pi\pi$ and $\sigma\sigma$ or \kk channels should be sizable.
Fits of similarly good quality were obtained with very 
different physical parameters in the \kk and \roro channels.
Lack of a sufficient number of observables and/or experimental precision, in
particular in the effective $4\pi$ channel, leads to the existence of several good
sets of model parameters with quite different channel and interchannel 
interactions.

We have studied positions of the $S$--matrix poles in the complex energy plane
($E = M- i\Gamma/2$). These positions have not been parameters of our fits.
In our model we have not included any kind of arbitrary background to fit the
data. We have only fitted the meson-meson interaction parameters and then 
analyzed the analytical structure of the reaction amplitudes.
 At low energy we find a very broad \epsig resonance 
(also called $\sigma$ meson) of a width between 510 and 560 MeV.
In the 3-channel fits the \fo resonance is seen in a vicinity of
the \kk threshold with a width of about 60 to 70 MeV.
The relatively narrow state \epw appears in the 3--channel fits. Its mass varies
 from about 1400 MeV to 1460 MeV and the width is about 150 MeV.
This width is close to values found by the 
Crystal Barrel Group (Ref. \cite{amslerklkl}), however the resonance 
masses obtained in our fits are smaller than their values of about 1500 MeV.

Finally, let us mention an important  difference between the fits
presented here for data 
taken on a polarized target in comparison with those performed with data 
\cite{grayer}  
obtained on a nonpolarized target.
If we fit the data of Ref. \cite{grayer} using our 3--channel model, we obtain a
very wide resonance at $M=1521$ MeV of width 503 MeV. This means that the recent 
analysis of data \cite{klr} supplies new information on the \epw meson.

In conclusion, we have analysed the isoscalar S--wave \pp and \kk scattering
using the CERN--Cracow--Munich data \cite{klr} 
in the framework of
the 2- and the 3--channel models of meson--meson scattering. 
All fits of the phase shifts analysis \cite{klr} indicate
presence of a relatively narrow scalar resonance of mass
1400 -- 1460 MeV. This resonance is quite compatible with recent 
observations of a possible scalar glueball at 1500 MeV with a width of 100 MeV.

\vskip 0.2cm    
This work has been performed in the framework of the IN2P3 -- Polish Laboratories
Convention (project No 93-71).

\vskip 1 cm

   \begin{figure}[ptb]

\vspace{-3cm}

\xslide{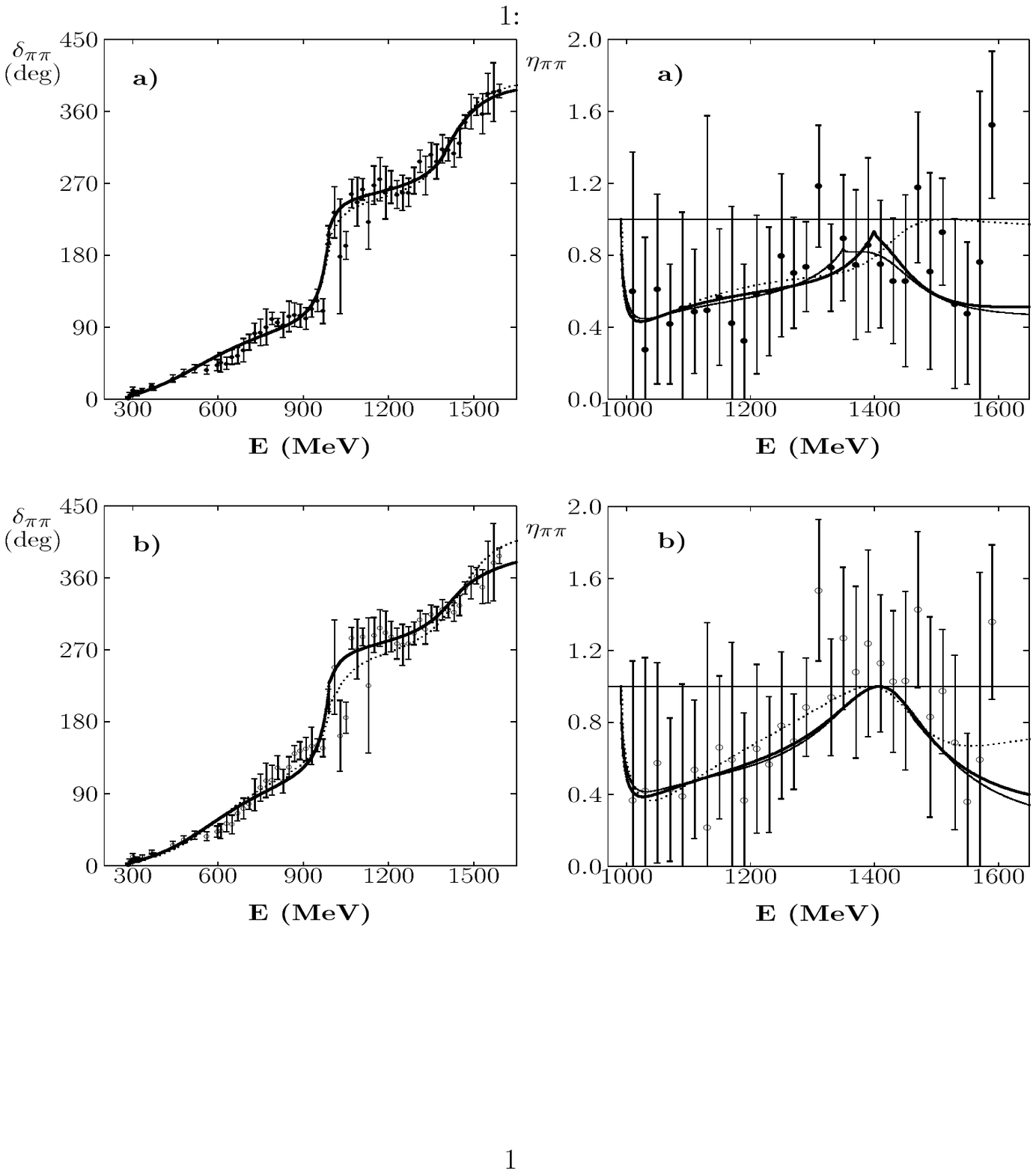}{15cm}{87}{245}{528}{626}{15.5cm} 

  \begin{centering}

  \caption[x]{ Energy dependence of $\pi\pi$ phase shifts and 
  inelasticities: 
a) fit to  "down-flat" data of ref. \cite{klr}~,
thick solid line corresponds to fit A, thin solid line to fit B and
dotted line to the 2--channel model fit;
b) fit to "up-flat" data of ref. \cite{klr}~, 
thick solid line corresponds to fit C, thin solid line to fit D and
dotted line to the 2--channel model fit.}

  \label{fig: fig1}

  \end{centering}

   \end{figure}



\thebibliography{References}

\bibitem{amslerklkl} A. Abele {\em et al.}, Phys. Lett. {\bf
B385} (1996) 425.
\bibitem{klm} R. Kami\'nski, L. Le\'sniak and J.-P. Maillet,
              Phys. Rev. {\bf D50}, 3145 (1994).
\bibitem{grayer} G. Grayer {\em et al.}, Nucl. Phys. {\bf B75}, 189 
(1974).
\bibitem{cohen} D. Cohen {\em et al.}, Phys. Rev. {\bf D22} (1980) 2595.
\bibitem{klr} R. Kami\'nski, L. Le\'sniak and K. Rybicki, 
              Z. Phys. {\bf C74}, 79 (1997).
\bibitem{kll} R. Kami\'nski, L. Le\'sniak and  B. Loiseau,
              Phys. Lett. {\bf B413}, 130 (1997).           
\bibitem {lles96} L. Le\'sniak, Acta Phys. Pol. {\bf B27}, 1835 (1996).
\end{document}